\documentclass[aps,pre,twocolumn,groupedaddress,amsmath,amssymb]{revtex4-1}
\makeatletter
\newcommand*{\rom}[1]{\expandafter\@slowromancap\romannumeral #1@}
\makeatother
\usepackage{graphicx}
\usepackage{dcolumn}
\usepackage{bm}
\usepackage{amsmath}
\usepackage{epstopdf}
\usepackage{amsfonts}
\usepackage{amssymb}
\usepackage[mathscr]{eucal}
\usepackage{hyperref}
\usepackage{xcolor}
\usepackage{float}
\usepackage{natbib}
\graphicspath{{figs/}}
\newcommand\prt{\partial}
\newcommand{\nn}{\nonumber}

\newcommand{\baa}{\begin{align}}
\newcommand{\eaa}{\end{align}}
\def\bse{\begin{subequations}} \def\ese{\end{subequations}}
\newcommand{\be}{\begin{equation}}
\newcommand{\ee}{\end{equation}}
\newcommand{\bea}{\begin{eqnarray}}
\newcommand{\eea}{\end{eqnarray}}

\begin{document}

\title{Note on WAdS$_3$ black holes in extended BHT gravity}
\author{Davood Mahdavian Yekta}
\email{d.mahdavian@hsu.ac.ir}
\affiliation{Department of Physics, Hakim Sabzevari University, P.O. Box 397, Sabzevar, Iran}
\pacs{03.65.Ca, 03.65.Sq, 04.70.Dy, 11.15.Yc}

\begin{abstract}

In this letter, we study a holographic diffeomorphism invariant higher-derivative extension of Bergshoeff-Hohm-Townsend (BHT) cosmological gravity in the context of Wald's formalism. We calculate the entropy, mass and angular momentum of warped anti-de Sitter (WAdS$_3$) black holes in ghost-free BHT massive gravity and its extension using the covariant phase space method. We also compute the central charges of the dual boundary conformal field theories (CFT) from the thermodynamics method.
\end{abstract}
\maketitle

\section{Introduction}\label{1}
Though the holographic nature of quantum gravity was originally suggested for black hole (BH) entropy \cite{Bekenstein:1973ur,Hawking:1974rv,Hawking:1975vcx}, the AdS/CFT correspondence \cite{Maldacena:1997re} provides the first explicit example of how holography can be realized in string theory. It is an important tool to learn about the physical properties of various quantum field theories in the strongly coupled non-perturbative regime, namely, perturbative string theory on asymptotically locally  AdS$_{d+1}$ backgrounds and conventional $d$-dimensional quantum field theories at their conformal fixed points \cite{Gubser:1998bc,Witten:1998qj}. It is well known that perturbative quantum gravity in four spacetime dimensions suffers from the problem of nonrenormalizability. This may be cured by going to lower dimensions. Three-dimensional (3D) gravity is an interesting playground to study problems in quantum gravity. 3D gravity with or without cosmological constant is described by a Chern-Simons (CS) gauge theory \cite{Deser:1983tn,Witten:1988hc}.

Though 3D gravity with a cosmological constant has no local degrees of freedom, there are global degrees of freedom, that is the theory admits Banados-Teitelboim-Zanelli (BTZ) BH solutions \cite{Banados:1992wn}. The first geometric modification of cosmological 3D gravity is topologically massive gravity (TMG) which includes an odd-parity gravitational CS term \cite{Deser:1981wh,Deser:1982vy}, while the BHT gravity is a higher-curvature parity-preserving modification (named also new massive gravity (NMG)) \cite{Bergshoeff:2009hq}. The latter theory describes a massive graviton with the same dynamics of a Fierz-Pauli theory at the linearized level, but demanding unitarity of the bulk theory spoils the unitarity of the boundary theory \cite{Bergshoeff:2009aq}. Other gravitational extensions of new massive gravity (ENMG) have been proposed in \cite{Sinha:2010ai,Gullu:2010pc,Paulos:2010ke} and with Maxwell field in \cite{Moussa:2008sj,Ghodsi:2010gk,Ghodsi:2010ev}. However, a bulk/boundary unitary theory in AdS$_3$ background has been introduced in \cite{Bergshoeff:2014pca} as minimal massive gravity (MMG) \cite{Bergshoeff:2014pca}. These models also admit WAdS$_3$ BH as another solution to the equations of motion \cite{Anninos:2008fx}.

In general, the conserved charges of BHs such as mass and angular momentum are expressed as integrals at spatial infinity; in the case of asymptotically-flat spacetimes they are denoted by Arnowitt-Deser-Misner (ADM) formula \cite{Arnowitt:1962hi} while for AdS backgrounds they are defined by extended Abbott-Deser-Tekin (ADT) method \cite{Abbott:1981ff,Deser:2002jk} in terms of an integral over the metric perturbation near spatial infinity. However, the ADT method does not yield satisfactory results for the higher curvature gravities like NMG or ENMG in the case of WAdS geometries. There is a background independent method proposed by Wald \cite{Wald:1993nt,Iyer:1994ys} to calculate the conserved charges using Noether charge and symplectic potential on the covariant phase space. One of our main interests is to calculate the conserved charges of WAdS$_3$ BHs in NMG and ENMG using Wald's formalism. Calculations for the BTZ BHs could be found in \cite{Nam:2016pfp,Nam:2018gju,Bergshoeff:2019rdb}.

According to AdS/CFT, one may expect that for each sector of 3D gravity with asymptotically AdS$_3$ geometry, there exists a dual 2D CFT. It would be of interest to study properties of 2D CFTs from different aspects such as BH thermodynamics, asymptotic symmetry group (ASG), CS formalism, and so on. From holographic point of view, the ASG of AdS$_3$ spaces is generated by two copies of the 2D conformal Virasoro algebra with central charges $c_R = c_L = {3\ell\over 2G}$, where $\ell$ is the AdS$_3$ radius and $G$ is the 3D Newton's constant \cite{Brown:1986nw}. The ASG has been applied for different 3D gravity theories in \cite{Liu:2009bk,Liu:2009kc,Ghodsi:2011ua} and for WAdS$_3$ backgrounds in \cite{Compere:2008cv,Compere:2009zj,Blagojevic:2009ek,MahdavianYekta:2016wwr}. However, it is not clear if the same correspondence could be set up in a theory with more general higher derivatives, such as ENMG for WAdS$_3$ BHs. This is our next motivation in this paper to compute the central charges from an approach rather than ASG.

In the context of BH holography, the central charges of dual CFTs can be computed from thermodynamics of the outer and inner horizons, the so-called thermodynamics method \cite{Chen:2012mh}. For a diffeomorphism-invariant theory of gravity, when the temperatures and entropies of outer and inner horizons satisfy $T^{+}\! S^{+}\!\!=\!\!T^{-}\! S^{-}$, then
\be \label{prod}
c_{L}=c_{R}=6\, \frac{d}{dN} \Big(\frac{S^{+}S^{-}}{4\pi^2}\Big),
\ee
where $N$ is the conserved charge of BH in the gravity sector \cite{Chen:2012yd,Chen:2013rb}. This methodology has been used for higher-dimensional BHs in \cite{Chen:2013rb,Golchin:2020mkh} and for 3D ones in \cite{Chen:2013aza,Zhang:2014wsa,MahdavianYekta:2020lde}. We will examine this conjecture for solutions of ENMG theory, particularly WAdS$_3$. In this regard, we need the physical conserved charges of the corresponding BHs which were cited in the first motivation.

In this letter, we initially review the first order formalism based on the covariant phase space method in sec. \ref{sec2} and derive integral expressions for the conserved charges independent of the background geometry. We also show taht the entropy is a conserved charge associated to vinishing Killing vector $\xi\!=\!{\prt_t}\!+\!\Omega_{\mathcal{H}}{\prt_\phi}$ on the balck hole horizon. In sec. \ref{sec3}, we compute the conserved charges for the WAdS$_3$ BH in NMG theory for the first time using this method. The generalization for both BTZ and WAdS$_3$ BHs in ENMG theory is extensively studied in sec. \ref{sec4} as well. We will obtain the central charges of the dual CFTs from thermodynamics method in sec. \ref{sec5} and show that our results are consisitent with literature. We conclude this letter in the last section.
\section{Conserved charges in Wald formalism}\label{sec2}
The first law of BH mechanics for stationary BHs can be derived from a diffeomorphism-invariant Lagrangian \cite{Wald:1993nt,Iyer:1994ys}.
For a BH with Killing vector $\xi$, vanishing on the Killing horizon, the conserved charges are well defined at spatial infinity. Also the BH entropy is the integral of the diffeomorphism Noether charge associated with $\xi$. The variation of $n$-form Lagrangian $L$ w.r.t the fields $\Phi$ is given by
\be\label{varL} \delta L=E_{\Phi}\delta \Phi+d \Theta(\Phi,\delta \Phi),\ee
where $E_{\Phi}$ is the field equations, i.e. $E_{\Phi}\!=\!0$ and $\Theta$ is the symplectic potential. The $(n-1)$-form symplectic current is
$ \omega(\Phi,\delta \Phi_1,\delta \Phi_2)\!=\!\delta_1 \Theta(\Phi,\delta_2 \Phi)\!-\!\delta_2 \Theta(\Phi,\delta_1 \Phi)$ and the symplectic form over a Cauchy surface $\Sigma$ is
\be\label{symp2} \Omega (\Phi,\delta \Phi_1,\delta \Phi_2)=\int_{\Sigma} \omega(\Phi,\delta \Phi_1,\delta \Phi_2).\ee

For a given $\xi$, the diffeomorphism invariance of $L$ implies $\delta_{\xi}L\!=\!\mathscr{L}_{\xi} L=d\imath_{\xi} L$ where $\imath_{\xi} L$ is the interior product $\xi\!\cdot\! L$ and $\delta_{\xi}\Phi\!\!=\!\!\mathscr{L}_{\xi} \Phi$ is the Lie derivative. We can associate a Noether current to each vector $\xi$ as $ J_{\xi}\!=\!\Theta(\Phi,\mathscr{L}_{\xi} \Phi)\!-\!\imath_{\xi} L$
such that from (\ref{varL}) we obtain
\be\label{jc} 
dJ_{\xi}=d\Theta(\Phi,\mathscr{L}_{\xi} \Phi)-d\imath_{\xi} L=-E_{\Phi} \mathscr{L}_{\xi} \Phi.
\ee
Whenever the equations of motion are satisfied, i.e. $E_{\Phi}\!=\!0$, $J_\xi$ is closed and
\be\label{NC2} J_{\xi}=dQ_{\xi},\ee
where $(n-2)$-form $Q_{\xi}$ is the Noether charge associated to $\xi$ and its integral over $\Sigma$ is called the ``\textit{conserved charge}''. 
The first law of BH mechanics can be deduced by considering the variation of $J_{\xi}$ with respect to the dynamical field $\delta \Phi$, i.e.
\be\label{DNC2} \delta J_{\xi}=\delta\Theta(\Phi,\mathscr{L}_{\xi} \Phi)-\imath_{\xi} \delta L\,,\ee
where from (\ref{varL}) and $E_{\Phi}\!=\!0$, the second term becomes
$ \imath_{\xi} \delta L\!=\!\mathscr{L}_{\xi}\Theta(\Phi,\delta\Phi)-d \imath_{\xi} \Theta(\Phi,\delta\Phi)$. Thus,
\be\label{DNC3} \delta J_{\xi}=\delta\Theta(\Phi,\mathscr{L}_{\xi} \Phi)-\mathscr{L}_{\xi}\Theta(\Phi,\delta \Phi)+ d \imath_{\xi} \Theta(\Phi,\delta \Phi)\,.\ee

The variation of Hamiltonian $H_\xi$ conjugate to $\xi$ is defined by $\delta H_{\xi}\!=\!\int_{\Sigma}\omega (\Phi,\delta \Phi,\mathscr{L}_{\xi}\Phi)$. From (\ref{DNC3}), one can obtain
$ \omega (\Phi,\!\delta \Phi,\mathscr{L}_{\xi}\Phi)\!=\!\delta J_{\xi}\!-\!d\imath_{\xi} \Theta(\Phi,\!\delta\Phi)$, then using (\ref{NC2})
\be\label{varH2}\delta H_{\xi}=\int_{\Sigma} \delta d Q_{\xi}-d\imath_{\xi} \Theta=\oint_{\prt \Sigma} \delta Q_{\xi}-\imath_{\xi}\Theta,\ee
where $\prt \Sigma$ is the boundary of Cauchy surface.
At the linearized level $\omega (\Phi,\delta \Phi,\mathscr{L}_{\xi}\Phi)=0$, which implies $\delta H_{\xi}=0$ and consequently
\be\oint_{\prt \Sigma} \delta Q_{\xi}-\imath_{\xi}\Theta=0.\ee
Suppose that $\xi$, a Killing vector vanishing on $\Sigma$, is defined by $\xi={\prt_t}+\Omega_{\mathcal{H}} {\prt_\phi}$ where $\Omega_{\mathcal{H}}$ denotes the angular velocity of the horizon and hereafter ${\prt_t}=\prt/{\prt t}$ and ${\prt_\phi}=\prt/{\prt\phi} $. Now, if $\xi$ vanishes on a bifurcation surface $\mathcal{H}$ and the hypersurface $\Sigma$ has its outer boundary at spatial infinity and interior boundary at $\mathcal{H}$, then
\be\label{CC} \int_{\mathcal{H}}\delta Q_{\xi}=\int_{\infty} \delta Q_{\xi}-\imath_{\xi}\Theta=\int_{\infty} \delta \chi_{\xi}.\ee
Nonetheless, if the theory admits a suitable definition of “canonical energy” $\mathcal{E}$ and “canonical angular momentum” $\mathcal{J}$, the variations of these quantities should be given by
\be\label{mj1} \delta\mathcal{E}=\int_{\infty} \delta \chi_{\xi}\left[\prt_t\right] ,\quad \delta\mathcal{J}=-\int_{\infty} \delta \chi_{\xi}\left[\prt_\phi\right]. \ee

Therefore, the variations of mass and angular momentum of a stationary rotating BH are defined by
\be\label{mj2} \delta\mathcal{M}\!=\!\frac{1}{8\pi G}\!\int_{\infty} \!\delta \chi_{\xi}\left[{\prt_t}\right] ,\quad \delta\mathcal{J}\!=\!-\frac{1}{8\pi G}\!\int_{\infty}\! \delta \chi_{\xi}\left[{\prt_\phi}\right]. \ee
Comparing Eqs. (\ref{CC}) and (\ref{mj1}) with the first law of thermodynamics, i.e. $T_{\mathcal{H}} \delta S=\delta {\mathcal{E}}-\Omega_{\mathcal{H}} \delta {\mathcal{J}},$
the entropy of BH is also given by
\be\label{ent} S=\frac{2\pi}{\kappa} \int_{\mathcal{H}} Q_{\xi},\ee
 where $\kappa$ is the surface gravity of the unperturbed BH. (For more details one can see e.g. \cite{Nam:2016pfp})
\section{New Massive Gravity}\label{sec3}
NMG theory is described by the following action \cite{Bergshoeff:2009hq}
\be\label{NMGa} \!\!\!\!\! I\!=\!\frac{1}{2\kappa}\!\int \! d^3 x\sqrt{-g} \left[\sigma R\!-\!2\Lambda_0\!-\!\frac{1}{m^2}\!\Big(R_{\mu\nu}R^{\mu\nu}\!-\!\frac38 R^2\Big)\right]\!\!,\ee
where $\kappa\!=\!8\pi G_3$, $\Lambda_0$ is a cosmological constant and $m$ is a mass parameter. Its CS Lagrangian is as follows \cite{Afshar:2014ffa}
 \be\label{Lnmg}\!\!\!\! L\!=\!-\sigma e\cdot R+\frac{\Lambda_0}{6}e\cdot e\times e-\frac{1}{m^2}\!\left[h\! \cdot \! De\!+\!f \!\cdot \!\big(R\!+\!\frac12 e\!\times\! f\big)\!\right]\!\!,\ee
where $\sigma\!=\!\pm 1$ is a sign parameter and $\{e^a, \omega^a, f^a, h^a\}$ are 1-form dynamical fields ($a=0,1,2$). $R\!=\!d\omega+\frac12 \omega\times\omega$ and $De\!=\!de+\omega\times e$ are curvature 2-form and Lorentz covariant torsion, respectively. The field equations are
\bse
\bea \label{eom1}
&-\sigma R+\frac{\Lambda_0}{2} e\times e-\frac{1}{m^2} \big(Dh+\frac12\,f\times f\big)=0,&\\
\label{eom10}&Df-m^2(e \times h)=0,&\\
\label{eom100}&R+e \times f=0,&\\
\label{eom1000}&De=0.&
\eea
\ese

The 2-form symplectic potential from (\ref{varL}) is given by
\be
\Theta=-\sigma \delta\omega \cdot e-\frac{1}{m^2} \left(\delta \omega \cdot f+\delta e \cdot h\right),\ee
hence, the Noether current is
\bea\label{NCnmg} J_{\xi}&=&dQ_{\xi}=\Theta(\Phi,\mathscr{L}_{\xi}\Phi)-i_{\xi} L\nn\\&=&d\left(-\sigma i_{\xi} \omega \cdot e-\frac{1}{m^2} \left(i_{\xi} \omega \cdot f+i_{\xi} e \cdot h\right)\right), \eea
and the corresponding Noether charge becomes
\be\label{NCnmg1} Q=-\sigma i_{\xi} \omega \cdot e-\frac{1}{m^2} \left(i_{\xi} \omega \cdot f+i_{\xi} e \cdot h\right).\ee
The variation of conserved charge from (\ref{CC}) yields
\bea\label{nmgkv}\delta \chi_{\xi}&\!\!\!=\!\!\!& -\sigma (i_{\xi} \omega \cdot \delta e+i_{\xi} e \cdot \delta \omega)\\
&\!\!\!-\!\!\!&\frac{1}{m^2} \left(i_{\xi} \omega \cdot \delta f+i_{\xi} f \cdot \delta\omega+i_{\xi} e \cdot \delta h+i_{\xi}  h \cdot \delta e\right).\nn\eea
For example, the mass and angular momentum of the BTZ BH have found in \cite{Nam:2016pfp} as
\be \label{mjbtz} M=\frac{r_{+}^2+r_{-}^2}{8G\ell^2}\Big(1-\frac{1}{\tilde{m}^2}\Big),\quad J=\frac{r_{+}r_{-}}{4G\ell}\Big(1-\frac{1}{\tilde{m}^2}\Big),\ee
hereafter $r_+ (r_-)$ refers to the outer (inner) horizon and $\tilde{m}^2\!=\!2m^2\ell^2$. However, the conserved charges of WAdS$_3$ BH have not been considered in \cite{Nam:2016pfp}. Thus, in the following subsection we will investigate this case.
\subsection{Conserved charges of WAdS$_3$ BH in NMG}

The WAdS$_3$ geometry is a fibration of real line over AdS$_{2}$ with a constant warped factor which is not asymptotically AdS$_3$ \cite{Moussa:2003fc,Anninos:2008fx,Clement:2009gq,Tonni:2010gb}. This breaks the $SL(2,R)_{L}\!\times\! SL(2,R)_{R}$ isometry group of AdS$_{3}$ down to $SL(2,R)\!\times\! U(1)$. Its BH metric is given by
\be\label{wads} \frac{ds^2}{\ell^2}=-F^2 dt^2+\frac{dr^2}{4F^2 K^2}+K^2 \left(d\phi+N dt\right)^2,\ee
where the metric functions $ F(r),\, K(r)$, and $N(r)$ are
\bea \label{func2}  \!\!\!\!F^2&\!=\!&\frac{(\nu^2\!+\!3)(r\!-\!r_{+})(r\!-\!r_{-})}{4 K^2},\,  N\!\!=\!\!\frac{2\nu r\!-\!\sqrt{(\nu^2\!+\!3) r_{+} r_{-}}}{2K^2},\nn\\
\!\!\!\!K^2&\!\!=\!&\frac{r}{4}\!\left[3(\nu^2\!-\!1)\!+\!(\nu^2\!+\!3)(r_{+}\!+\! r_{-})\!-\!4\nu \sqrt{(\nu^2\!+\!3) r_{+} r_{-}}\right]\!.\nn\eea
$\nu$ is a dimensionless parameter in the warped factor. The metric (\ref{wads}) is a solution of Eqs. (\ref{eom1})-(\ref{eom1000}) if 
\be \nu^2=\frac{\tilde{m}^2+3}{20},\quad\Lambda_0=\frac{\tilde{m}^4-234 \,\tilde{m}^2+189}{200\, \tilde{m}^2 \ell^2}.\ee
The 1-form frame fields and spin connections compatible with (\ref{wads}) are given as follows
\bea \label{ffwads}\begin{split}
&e^{0}=\ell F\,dt,\,\,\, e^1=\frac{\ell\, dr}{2FK},\,\,\, e^2=\ell K (d\phi+N\,dt),&\\
 \label{sc}& \omega^{0}\!=\!W_1 e^0\!+\!W_2 e^2\!,\,\,\, \omega^1\!=\!W_3 e^1\!,\,\,\, \omega^2\!=\!W_4 e^2\!+\!W_5 e^0\!,\,\,\,&\end{split}
\eea
where the functions $W_i$  ($i=1,...,5$) are
\be\label{scc} \!\!\!\!\!\! W_1\!=\!W_3\!=\!-W_4\!=\!\frac{K^2N'}{\ell},\,\, W_2\!=\!\!\frac{2F K'}{\ell}, \,\, W_5\!=\!\!\frac{2K F'}{\ell}.
\ee
The auxiliary fields $f^a$ and $h^a$ are found from Eqs. (\ref{eom100}) and (\ref{eom10}) as
\bea \label{f}\begin{split}&   \!\!\!\!\!\! 
f^{0}\!=\!U_1 e^0+U_2 e^2,\,\, f^1\!=\!U_3 e^1,\,\, f^2\!=\!U_4 e^2+U_5 e^0,&\\ \label{h}&  \!\!\!\!\!\!  h^{0}\!=\!V_1e^0+V_2 e^2,\,\, h^1\!=\!V_3 e^1,\,\, h^2\!=\!V_4 e^2+V_5 e^0,&\end{split}
\eea
with some lengthy exppressions for coefficient functions $U_i$ and $V_i$ that are given in the supplementary material for this letter. To calculate the charge variations (\ref{mj2}) we need the interior products. For variation ${\prt_t}$ we have
\bea\label{ip}\begin{split} &\imath_{\xi} e^0=\ell F\,,\quad \imath_{\xi} e^2=\ell K N,&\\&
\imath_{\xi}\omega^0=\ell F W_1\!+\!\ell K N W_2\,,\,\, \imath_{\xi} \omega^2\!=\!\ell K N W_4\!+\!F W_5,&\\&
\imath_{\xi}f^0=\ell F U_1\!+\!\ell K N U_2,\,\, \imath_{\xi} f^2\!=\!\ell K N U_4\!+\!F U_5,&\\&
\imath_{\xi}h^0=\ell F V_1\!+\!\ell K N V_2,\,\, \imath_{\xi} h^2\!=\!\ell K N V_4\!+\!F V_5,&\end{split}
\eea
and those for ${\prt_\phi}$ are
\bea\label{ipw2} \begin{split}
& \imath_{\xi} e^{2}=\ell K,\,\, \imath_{\xi} \omega^0=\ell KW_2,\,\, \imath_{\xi} \omega^2=\ell K W_4,&\\
& \imath_{\xi} f^0=\ell K U_2\,,\quad \imath_{\xi} f^2=\ell K U_4,&\\
& \imath_{\xi} h^0=\ell K V_2\,,\quad \imath_{\xi} h^2=\ell K V_4.&\end{split}
\eea

For variations at spatial boundaries on the Cauchy surface $\Sigma$, we only need to consider $d\phi$ component, i.e., 
\bea \label{varsc}\begin{split}
&\delta e^0=0,\quad \delta e^2=\ell \delta(K)d\phi\,,&\\
& \delta\omega^0=\ell\delta (K W_2) d\phi,\quad \delta\omega^2=\ell\delta(K W_4) d\phi\,,&\\
&\delta f^0\!=\!\ell\delta (K U_2) d\phi,\quad \delta f^2\!=\!\ell\delta(K U_4) d\phi,&\\
&\delta h^0\!=\!\ell\delta (K V_2) d\phi,\quad \delta h^2\!=\!\ell\delta(K V_4) d\phi.&
\end{split}
\eea
Then, the $\delta$ variations (\ref{nmgkv}) become
\bea\label{varkv30}
&&\!\!\!\!\delta \chi_{\xi}\left[{\prt_t}\right] \!=\!\alpha\,\nu \Big(\!(\delta r_{+}\!+\!\delta r_{-})\!-\!\tfrac{(\nu^2 +3)}{2\nu}\tfrac{(r_{+} \delta r_{-}\!+r_{-} \delta r_{+})}{\sqrt{r_{+} r_{-} (\nu^2\! +\!3)}}\!\Big)d\phi,\nn\\
&&\!\!\!\!\delta \chi_{\xi}\left[\prt_\phi\right]\!=\!\alpha\,\nu^2\Big(\tfrac{(5\nu^2\!+\!3)}{\nu^2}(r_{-}\delta r_{+}\!+\!r_{+}\delta r_{-})\nn\\
&&\quad-\tfrac{(\nu^2 +3)}{4\nu}\tfrac{(3 r_{-}r_{+}+r_{-}^2) \delta r_{+}+(3 r_{-}r_{+}+r_{+}^2) \delta r_{-}}{\sqrt{r_{+} r_{-} (\nu^2 +3)}}\Big)\, d\phi\,,
\eea
where $\alpha\!=\!\frac{4\nu \ell(\nu^2+3)}{(20\nu^2-3)}$. Integrating (\ref{varkv30}) we obtain the mass and angular momentum of WAdS$_3$ BH as
\bea \label{mjwads} \!\!\!M&\!=\!&\tfrac{\alpha}{4G\ell}\Big(\nu(r_{+}+ r_{-})-\sqrt{(\nu^2+3) r_{+} r_{-}}\Big)\!,\\
\!\!\! J&\!=\!&\tfrac{\alpha}{16G}\!\Big[\!\Big(\!\nu(r_{+}\!+\! r_{-})\!-\!\sqrt{(\nu^2\!+\!3) r_{+} r_{-}}\Big)^2\!-\!\nu^2 (r_{+}\!-\! r_{-})^2\!\Big]\!,\nn\eea
which are compatible with results from Clement's approach \cite{Clement:1994sb} in \cite{Ghodsi:2011ua,Tonni:2010gb} and canonical formalism in \cite{MahdavianYekta:2016wwr}.
\section{Extended New Massive Gravity}\label{sec4}
ENMG is a higher order curvature deformation of BHT gravity from holographic $c$-theorem in the context of AdS/CFT correspondence \cite{Sinha:2010ai}. This theory is described by the following action
\bea\label{ENMGa} I&\!=\!&\frac{1}{2\kappa}\int d^3 x\,\sqrt{-g} \Big[\sigma R-2\Lambda_0-\frac{1}{m^2}\Big(R_{\mu\nu}R^{\mu\nu}-\frac38 R^2\Big)\nn\\
&\!\!\!-\!\!\!&\frac{2}{3m^4}\Big(R_{\mu}^{\nu}R_{\nu}^{\rho}R_{\rho}^{\mu}-\frac98 RR_{\mu\nu}R^{\mu\nu}+\frac{17}{64}R^3 \Big)\Big].\eea
This action can also be found in the infinitesimal curvature expansion of a Born-Infeld-like action\cite{Gullu:2010pc}
\bea\label{BINMG0}
S\!=\!\frac{2m^2}{\kappa}\!\int \!d^3 x\Big[\sqrt{-|g_{\mu\nu}\!+\!\frac{\sigma}{m^2}\!G_{\mu\nu}|}\!-\!(1\!+\!\frac{\Lambda}{2m^2})\sqrt{-g}\Big]\!,\nn
\eea
up to the corresponding order. The CS Lagrangian in the methodology of extended gravities \cite{Afshar:2014ffa} is given by
\bea\label{Lenmg} \!\!\!\! L&\!=\!&-\sigma\, e\cdot R-\frac{\Lambda_0}{6}\,e\cdot e\times e+\frac{1}{2m^2} e\cdot f\times f\nn\\
&&-\frac{1}{m^4}\left[h \cdot De-\frac16 f\cdot f\times f+k \cdot \left(R+e\times f\right)\right]\!,\eea
where $k^a$ is a new auxiliary field in addition to the previous set. The ENMG fields equations are
\bse\bea\label{eom2}
&\!\!\!\!\!\!\!\!\!\!\!-\sigma R-\frac{\Lambda_0}{2} e\times e\!+\!\frac{1}{2m^2}  f\times f\!-\!\frac{1}{m^4} \big(Dh\!+\!f \!\times\! k\big)\!=\!0, &\\
\label{eom21}&Dk+e \times h+\sigma m^4  De=0,&\\
\label{eom22}&e \times k-\frac12 f \times f-m^2 e \times f=0,&\\
\label{eom23}&R+e \times f=0,&\\
\label{eom24}& De=0,&
\eea\ese
and the symplectic potential is as follows
\be\Theta=-\sigma \delta\omega \cdot e-\tfrac{1}{m^4} \left(\delta e \cdot h+\delta \omega \cdot k\right).\ee
Then, the Noether current becomes
\be\label{NC3} J_{\xi}=d\Big(-\sigma i_{\xi} \omega \cdot e-\tfrac{1}{m^4} \left(i_{\xi} e \cdot h+i_{\xi} \omega \cdot k\right)\Big), \ee
which is associated to the Noether charge $Q_{\xi}=-\sigma i_{\xi} \omega \cdot e-\frac{1}{m^4} \left(i_{\xi} e \cdot h+i_{\xi} \omega \cdot k\right)$.
Therefore, the charge variation (\ref{CC}) corresponding to Killing vector $\xi$ yields
\bea\label{varkv}\!\!\!\!\!\!\delta \chi_{\xi}&\!=\!&-\sigma (i_{\xi} \omega \cdot \delta e+i_{\xi} e \cdot \delta \omega)\nn\\
&&-\frac{1}{m^4}\! \left(i_{\xi} e \cdot \delta h\!+\!i_{\xi}  h \cdot \delta e\!+\!i_{\xi} \omega \cdot \delta k\!+\!i_{\xi} k \cdot \delta\omega\right)\!.\eea
In the rest of this section we compute the mass and angular momentum of both the BTZ and WAdS$_3$ BHs from (\ref{varkv}) which have not been considered in literature for this formalism.
\subsection{Conserved charges of BTZ BH in ENMG}
The most well-known locally AdS$_3$ solution of all 3D gravities is the BTZ BH \cite{Banados:1992wn} with line element
\be\label{BTZ} ds^2=-F^2 dt^2+F^{-2}{dr^2}+r^2 \left(d\phi+N dt\right)^2,\ee
where the metric functions $F(r)$ and $N(r)$ are
\be \label{func1} F^2\!=\!{(r^2\!-\!r_{+}^2)(r^2\!-\!r_{-}^2)}/{r^2 \ell^2},\quad N\!=\!-{r_{+} r_{-}}/{\ell\, r^2 },
\ee
and $\ell$ is the radius of AdS$_3$ space. From (\ref{BTZ}) and Eq. (\ref{eom24}) the frame fileds and spin connections are
\bea
&\label{ffbtz} e^{0}=F\,dt\,,\quad e^1=\frac{dr}{F}\,,\quad e^2=r(d\phi+N\,dt),&\\
&\!\!\!\!\omega^{0}\!\!=\!\!\frac12 r N' e^0\!+\!\frac{1}{r}F e^2,\,\, \omega^1\!=\!\frac12 r N' e^1,\,\, \omega^2\!=\!F' e^0\!-\!\frac12 r N' e^2\!.&\nn
\eea
Using these fields and Eqs. (\ref{eom21})-(\ref{eom23}), we obtain
\bea
&\label{forms} f^{a}=\frac{1}{2\ell^2}\, e^{a},\quad k^a=\big(\frac{1}{8\ell^2}+\frac{m^2}{2\ell^2}\big) e^{a},\quad h^{a}=0.&
\eea
For $\sigma\!=\!-1$ and from Eq. (\ref{eom2}), the cosmological term is $\Lambda_0\!=\!-\frac{2\tilde{m}^4+\tilde{m}^2+1}{2\tilde{m}^4 \ell^2}<0$,
which again confirms that the BTZ BH is an AdS$_3$ geometry.

The variation (\ref{varkv}) is as follows
\be\label{varkv2} \delta \chi_{\xi}=-\big(\sigma+\frac{1}{\tilde{m}^2}+\frac{1}{2\tilde{m}^4}\big) (i_{\xi} \omega \cdot \delta e+i_{\xi} e \cdot \delta \omega),\ee
and the interior products for ${\prt_t}$ and ${\prt_\phi}$ are given by
\bea
\label{ipew}&\!\!\!\!\!\! \imath_{\xi} e^0\!\!=\!\!F\!,\, \imath_{\xi} e^2\!\!=\!\!r N\!, \imath_{\xi} \omega^0\!\!=\!\!F(N\!+\!\tfrac12 rN'\!)\!, \imath_{\xi} \omega^2\!\!=\!\!F F'\!\!-\!\tfrac12 r^2 N N'\!,&\nn\\
 &\imath_{\xi} e^{2}=r,\quad \imath_{\xi} \omega^0=F,\quad \imath_{\xi} \omega^2=-\frac12 r^2 N'\,.&
\eea
The variations related to $d\phi$ are given by $\delta\omega^0\!=\!\delta F d\phi$ and $\delta\omega^2\!=\!-\frac12 r^2 \delta N' d\phi$. Thus, from (\ref{varkv2}) and for $\sigma\!=\!-1$ we obtain
\bea\label{varkv3} \delta\chi_{\xi}\!\left[{\prt_t}\right] &\!\!=\!\!&\big(1\!-\!\frac{1}{\tilde{m}^2}\!-\!\frac{1}{2\tilde{m}^4}\big)(r_{+} \delta r_{+}\!+\!r_{-} \delta r_{-})\, d\phi,\\
\delta\chi_{\xi}\!\left[{\prt_\phi}\right]&\!\!=\!\!&-\big(1\!-\!\frac{1}{\tilde{m}^2}\!-\!\frac{1}{2\tilde{m}^4}\big)(r_{+} \delta r_{-}\!+\!r_{-} \delta r_{+})\, d\phi.\nn
\eea
Substituting these phrases in (\ref{mj2}) and integrating we find
\bea \label{mjbtz}M&\!=\!&\frac{r_{+}^2+r_{-}^2}{8G\ell^2}\Big(1\!-\!\frac{1}{\tilde{m}^2}\!-\!\frac{1}{2\tilde{m}^4}\Big),\\ J&\!=\!&\frac{r_{+}r_{-}}{4G\ell}\Big(1\!-\!\frac{1}{\tilde{m}^2}\!-\!\frac{1}{2\tilde{m}^4}\Big),\nn\eea
which are also consistent with calculations in \cite{Ghodsi:2011ua,Tonni:2010gb}.
\subsection{Conserved charges of WAdS$_3$ BH in ENMG}
The WAdS$_3$ BH (\ref{wads}) is also a solution of the field equations (\ref{eom2})-(\ref{eom24}) if 
\bea\label{p2} 
&\nu^2=\frac67+\frac{5\tilde{m}^2}{21}+\frac{\Delta}{42},&\\
&\!\!\!\!\!\!\!\!\!\!\Lambda_0\!=\!\tfrac{(148 \tilde{m}^4\!-270 \tilde{m}^2\!+1215)\Delta+1081 \tilde{m}^6\!+14238 \tilde{m}^4\!-6075 \tilde{m}^2\!-32805}{9261 \tilde{m}^4 \ell^2},&\nn
\eea
where $\Delta\!=\!\sqrt{58 \tilde{m}^4 \!+\!594 \tilde{m}^2\!+\!729}$. The fields $e^a$, $\omega^a$, and $f^{a}$ are given respectively by (\ref{ffwads}) and first row of (\ref{f}), while the auxiliary fields $k^{a}$ and $h^a$ from Eqs. (\ref{eom22}) and (\ref{eom21}) are
\bea \begin{split}&\label{kwads} \!\!\! k^{0}\!=\!P_1 e^0\!+\!P_2\,e^2,\,\,\, k^{1}\!=\! P_3\, e^1,\,\,\, k^{2}\!=\! P_4\, e^2\!+\! P_5\,e^0,&\\
&\label{hwads} \!\!\! h^{0}\!=\!Y_1 e^0\!+\! Y_2\,e^2,\,\,\, h^{1}\!=\! Y_3 e^1,\,\,\, h^{2}\!=\! Y_4 e^2\!+\!Y_5\,e^0,&\end{split}
\eea
where again the lengthy coefficients $P_i$ and $Y_i$ are given in the supplementary material. 

The interior products for ${\prt_t}$ are
\bea \label{ipk}\begin{split} &
\imath_{\xi}k^0\!=\!\ell F P_1\!+\!\ell k N P_2,\,\,\, \imath_{\xi} k^2\!=\!\ell K N P_4\!+\!F P_5,&\\
&\imath_{\xi}h^0\!=\!\ell F Y_1\!+\!\ell k N Y_2,\,\,\, \imath_{\xi} h^2\!=\!\ell K N Y_4\!+\!F Y_5,&\end{split}
\eea
and for ${\prt_\phi}$
\bea 
\label{ipw2}\begin{split}
 & \imath_{\xi} k^0=\ell K P_2\,,\quad \imath_{\xi} k^2=\ell K P_4\,,&\\
 & \imath_{\xi} h^0=\ell K Y_2\,,\quad \imath_{\xi} h^2=\ell K Y_4\,.& \end{split}
\eea
The variations of the form fields with respect to $d\phi$ are given by the first row of (\ref{varsc}) and
\bea\label{varsc2}\begin{split}
&\delta k^0=\ell\delta (K P_2) d\phi,\quad \delta k^2=\ell\delta(K P_4) d\phi,&\\
& \delta h^0=\ell\delta (K Y_2) d\phi,\quad \delta h^2=\ell\delta(K Y_4) d\phi.&\end{split}\eea

Now, substituting the above relations in the charge variation (\ref{mj2}), one can obtain
\bea\label{vkenmg}
\delta \chi_{\xi}\left[{\prt_t}\right] &\!=\!&4\nu\Xi \Big((\delta r_{+}+\delta r_{-})\nn\\&\!\!\!-\!\!\!&\frac{(\nu^2 +3)}{2\nu}\frac{(r_{+} \delta r_{-}+r_{-} \delta r_{+})}{\sqrt{r_{+} r_{-} (\nu^2 +3)}}\Big)\,d\phi,\\
\delta \chi_{\xi}\left[{\prt_\phi}\right]&\!=\!&\nu^2\Xi\Big(\frac{(5\nu^2+3)}{\nu^2}(r_{-}\delta r_{+}+r_{+}\delta r_{-})\nn\\&\!\!\!-\!\!\!&\frac{(\nu^2\! +\!3)}{4\nu}\frac{(3 r_{-}r_{+}\!+\!r_{-}^2) \delta r_{+}\!+\!(3 r_{-}r_{+}\!+\!r_{+}^2) \delta r_{-}}{\sqrt{r_{+} r_{-} (\nu^2 +3)}}\Big) d\phi,\nn
\eea
 where
 \be\label{xi}\Xi\!=\!\frac{2\nu \ell (\nu^2+3) \big(16\nu^2+3-\sqrt{232 \nu^4+168\nu^2-45}\big)}{\left(20\nu^2-3-\sqrt{232 \nu^4+168\nu^2-45}\right)^2}.\nn\ee
Therefore, the ADM mass and angular momentum of WAdS$_3$ black hole in ENMG theory become
\bea \label{mjenmg} \!\!\!\!\!&M\!=\!\frac{\Xi}{G \ell}\big(\nu(r_{+}\!+\! r_{-})\!-\!\sqrt{(\nu^2+3) r_{+} r_{-}}\big),&\\
\!\!\!\!\! &J\!=\!\frac{\Xi}{4G}\!\Big[\!\big(\nu(r_{+}\!+\! r_{-}\!)\!-\!\sqrt{(\nu^2\!+\!3) r_{+} r_{-}}\big)^2\!-\!\nu^2 (r_{+}\!-\! r_{-})^2\Big]\!,\nn&\eea
The results are again consistent with calculations in \cite{Nam:2010dd}.
\section{Thermodynamics method} \label{sec5}
One can consider the inner Cauchy horizon of the charged or rotating BHs as a dynamical surface with intensive thermodynamical quantities \cite{Ansorg:2009yi,Cvetic:2010mn,Castro:2012av,Detournay:2012ug}. It is shown that not only the product of the outer and inner horizons is a mass-independent quantity but it also may have holographic application \cite{Larsen:1997ge,Cvetic:1997uw,Golchin:2019hlg}. 
In fact, the mass-independence of entropy product is an indication for a BH to have holographic description, but neither a sufficient nor a necessary condition. 
This product is universal if the thermodynamics satisfy the condition $T^+ S^+ \!=\! T^- S^-$ which happens generally for a class of diffeomorphism-invariant Lagrangians such as NMG or ENMG. The equivalency of central charges in Eq.~(\ref{prod}) can also be deduced from holographic c-theorem in diffeomorphism-invariant Lagrangians of gravity \cite{Kraus:2005vz}. 
On the other hand, though the BTZ and WAdS$_3$ BHs have two 2D CFT duals in TMG theory, the left- and right-moving central charges are different $c_L\!\neq\! c_R$ which is due to the holographic gravitational anomaly \cite{Kraus:2005zm}. Of course, it can also be deduced by the fact that the mass-independence of the entropy product breaks down or $T^+ S^+ \!\neq\! T^- S^-$ \cite{MahdavianYekta:2016kqh}.

In the context of BH/CFT, one can use the thermodynamics method to discuss the CFT duals \cite{Chen:2012mh,Chen:2012yd,Chen:2013rb}. Here, we use (\ref{prod}) to obtain the central charges of CFT duals for the BTZ and WAdS$_3$ BHs in ENMG theory. We have recently considered the case of WAdS$_3$ BH for NMG theory in \cite{MahdavianYekta:2020lde}.

\subsection{CFT dual of BTZ BH in ENMG}
Using the killing vector $\xi=\prt_t+\Omega_{\pm} \prt_{\phi}$ in Eq. (\ref{ent}), the entropies $S^{\pm}$ of outer and inner horizons are
\be\label{ebnmg}  S^{\pm}=\frac{\pi r_{\pm}}{2G}\Big(1-\frac{1}{\tilde{m}^2}-\frac{1}{2\tilde{m}^4}\Big),\ee
and the corresponding Hawking temperatures and angular velocities are defined by
\be\label{tbtz1} \!\!\!\! T^{\pm}\!=\!\frac{\kappa}{2\pi}\Big|_{r=r_{\pm}}\!=\!\frac{r_{+}^2\!-\!r_{-}^2}{2\pi\ell^2 r_{\pm}},\,\,\,\, \Omega^{\pm}\!=\!N^{\varphi}\Big|_{r=r_{\pm}}\!=\!\frac{r_{\mp}}{\ell r_{\pm}}.\ee
It is straightforward to check that the above quantities and (\ref{mjbtz}) satisfy the first law of BH thermodynamics and Smarr-like formula as $ dM=\pm T^{\pm} dS^{\pm}+\Omega^{\pm} dJ$ and $M=\pm \frac12 T^{\pm} S^{\pm}+\Omega^{\pm} J$.
Noting the relations (\ref{ebnmg}) and (\ref{tbtz1}) one has $T^+S^+=T^-S^-$, thus the entropy product is mass independent, i.e.,
\be\label{prod10}
S^{+}S^{-}=\frac{\pi^2 \ell}{G}J\Big(1-\frac{1}{\tilde{m}^2}-\frac{1}{2\tilde{m}^4}\Big).
\ee

According to (\ref{prod}) and by choosing $N=J$, the central charges of the dual CFT for the BTZ solution are
\be \label{CC1}
c_{L}\!=\!c_{R}\!=\!6\, \frac{d}{dJ} \Big(\frac{S^{+}S^{-}}{4\pi^2}\Big)\!=\!\frac{3\ell}{2G}\Big(1-\frac{1}{\tilde{m}^2}-\frac{1}{2\tilde{m}^4}\Big)\!,
\ee
which are compatible with the results of ASG group analysis in \cite{Liu:2009kc,Ghodsi:2011ua,MahdavianYekta:2016kqh} and holographic c-theorem \cite{Kraus:2005vz,Sinha:2010ai}. It is shown that the entropies (\ref{ebnmg}) and central charges (\ref{CC1}) satisfy the Cardy formula $ S^{\pm}=\frac{\pi^2}{3}(c_{L} T_{L}\pm c_{R}T_{R})$ as well, where $T_{L,R}=\frac{r_{+}\pm r_{-}}{2\pi\ell}$\, are given in \cite{Maldacena:1998bw}. Also the energies of the left- and right-moving sectors of the dual CFT are
\be\label{en1}\!\!\! E_{L}\!=\!\frac{\pi^2}{3}c_{L} T_{L}^2\!=\!\frac{M\ell+J}{2},\,\,\,\, E_{R}\!=\!\frac{\pi^2}{3}c_{R} T_{R}^2\!=\!\frac{M\ell\!-\!J}{2}.\ee
From these relations we have
\be M=(E_{R}+E_{L})/\ell\,,\qquad J=(E_{R}-E_{L})\,,\ee
 and using the central charges in Eq.~(\ref{CC1}), one can recast the entropy product as
\be  \frac{S^{+} S^{-}}{4\pi^2}\!=\!\frac{\ell}{12}\,\Big( (c_{R}\!+\!c_{L})\,( E_{R}\!-\!E_{L})\!+\!(c_{R}\!-\!c_{L})\,( E_{R}\!+\!E_{L})\Big),\ee
which explicitly shows that when $c_{L}\!=\!c_{R}$ the product $S_+S_-$ is mass-independent.
\subsection{CFT dual of WAdS$_3$ BH in ENMG}
Similar to the previous subsection, the entropies $S^{\pm}$ are computed from \ref{ent} as
\be \label{ewads}
S^{\pm}=\frac{16\pi\nu^2 \ell \Xi}{G} \Big(2\nu r_{\pm}-\sqrt{(\nu^2+3) r_{+} r_{-}}\Big),
\ee
and other thermodynamic quantities are given by
\bea\label{twads} T^{\pm}&=&\frac{(\nu^2+3)}{4\pi\ell} \frac{r_{+}- r_{-}}{\left(2\nu r_{\pm}-\sqrt{(\nu^2+3) r_{+} r_{-}}\right)},\nn\\ \Omega^{\pm}&=&\frac{2}{\ell \left(2\nu r_{\pm}-\sqrt{(\nu^2+3) r_{+} r_{-}}\right)}\,.\eea
Again (\ref{mjenmg}), (\ref{ewads}) and (\ref{twads}) satisfy $dM\!=\!\pm T^{\pm} dS^{\pm}\!+\!\Omega^{\pm} dJ$ and $ M\!=\!\pm T^{\pm} S^{\pm}\!+\!2\Omega^{\pm} J$. Since  $T^+S^+\!=\!T^-S^-$, the entropy product is universal, $S^{+}S^{-}\!=\!\frac{2(8\pi)^2\nu^3\ell \,\Xi}{G} J$, and the central charges of the dual CFT are
\be \label{CC2} c_{L}=c_{R}=\frac{192\nu^3\ell \,\Xi}{G}\,.
\ee
Following \cite{Anninos:2008fx} for the left and right temperatures $T_{L}\!\!=\!\!\frac{(\nu^2\!+\!3)}{8\pi \nu \ell} \big(\nu(r_{+}\!+\! r_{-})\!-\!\sqrt{(\nu^2\!+\!3) r_{+} r_{-}}\big)$ and $T_{R}\!\!=\!\!\frac{(\nu^2\!+\!3)}{8\pi \ell} (r_{+}\!-\! r_{-})$, it is possible to find the energies of the left- and right-moving sectors of the dual CFT from central charges (\ref{CC2}) and these temperatures. However, the mass and angular momentum of WAdS$_3$ BH can be represented vs. these energies as
\be\label{en2} M=\frac{1}{2\sqrt{2} \nu} \sqrt{\frac{ c_L{} E_{L}}{6}}, \quad J=\frac{\ell^2}{2(\nu^2+3)} (E_{L}-E_{R})\,.
\ee
The relations are very similar to ones obtained in \cite{Anninos:2008fx} for TMG and in \cite{Nam:2018gju} for MMG theory but are different in the coefficients due to ENMG deformation.
\section{Conclusion}
Our aim in this letter was devoted to investigate two related concepts for WAdS$_3$ and BTZ BHs in two diffeomorphism-invariant theories in 3D gravity, the so-called NMG and ENMG. In the former, we compute the masses and angular momenta of these BHs by employing the first order Wald formalism so that the variations of these charges  are satisfied with the first law of the BH thermodynamics. In the latter, we compute the central charges of the CFT duals to these gravity models through the thermodynamics method of BH/CFT correspondence. In fact without knowing the explicit forms of BHs, the thermodynamics method could give consistent results with the ones obtained from ASG analysis or Clement's approach.

In the context of CS formalism, we have described these 3D gravities with the Lagrangians consisting of frame fields, spin connections and some extra auxiliary fields. The conserved charges were calculated by integrating some charge variations $\delta \chi_{\xi}$ over the boundary of Cauchy surface which are defined in terms of some symplectic potential $\Theta$ and conserved Noether charge $Q_{\xi}$. In addition, we computed the entropy of BHs in Wald formalism by integrating the Noether charge over a biforcation horizon $\mathcal{H}$ with killing vector $\xi\!=\!\prt_t\!+\!\Omega_{\mathcal{H}} \prt_{\phi}$. 

Beside the quantities on the event horizon, we observed that thermodynamic quantities on the inner Cauchy horizon of all solutions satisfy the first law and Smarr relation as well. With explicit computations we showed that the product of BH entropies of the inner and outer horizons is independent of the mass which is compatible with the universality of diffeomorphism-invariant theories. On the other hand, we naively found that the thermodynamic quantities fulfill the condition $T^+S^+=T^-S^-$ and therefore using the thermodynamic method lead to equal right- and left-moving central charges of CFT duals for BTZ and WAdS$_3$ BHs in ENMG theories.
 \section*{Data Availability Statement}
 No new data were created or analysed in this study.


\bibliography{refs}

\newpage

\onecolumngrid

\begin{appendix}

\begin{center}
	\textbf{{\large Supplementary material for\\ ``Note on WAdS$_3$ black holes in extended BHT gravity''}}
\end{center}

In this supplementary material we refer to the exact expressions for the coefficient functions that are appeared in the definitions of auxiliary fields in the new massive gravity (NMG) and its extended (ENMG) theories for WAdS$_3$ black holes in the main text. The Larangians of both theories in the Chern-Simons formalism are taken from \cite{Afshar:2014ffa} while the methology of calculations are borrowed from \cite{Nam:2016pfp}.
\section{NMG functions}
The auxiliary fields $f^a$ and $h^a$ ($a=0,1,2$) in the equations of motion of NMG are defined as follows
 \bea \label{fh}
f^{0}=U_1 e^0+U_2 e^2,\,\, f^1=U_3 e^1,\,\, f^2=U_4 e^2+U_5 e^0,\qquad h^{0}=V_1e^0+V_2 e^2,\,\, h^1=V_3 e^1,\,\, h^2=V_4 e^2+V_5 e^0,
\eea
where the functions $U_i$ and $V_i$ ($i=1,2,...,5$) are given by
\bea\label{fcomps}
U_1&\!\!=\!\!&-\tfrac{2}{\ell^2}\Big(\tfrac{3(K^2 N')^2}{4} \!+\!(K K')' F^2\!-\!KK'FF'\!-\!K^2( F F''\!+\! F'^2)\Big)\!,\,\,U_2\!=\!\!-U_5\!=\!-\tfrac{2}{\ell^2}\! \left(K^2 F (4 K' N'\!+\!K N'')\right),\nn\\
U_3&\!\!=\!\!&-\tfrac{2}{\ell^2}\Big(\tfrac{3(K^2 N')^2}{4}-(K K')' F^2-KK'FF'-K^2( F F''+ F'^2)\Big),\\
U_4&\!\!=\!\!&\tfrac{2}{\ell^2}\Big(\tfrac{5(K^2 N')^2}{4}+(K K')' F^2+KK'FF'-K^2 ( F F''+ F'^2)\Big),\nn
\eea
and
\bea \label{hcomps}
V_1&\!=\!&-\tfrac{4 K^2}{m^2 \ell^3}\Big(\!F^2(12 K'^2 N'\!+\!8K K' N''\!+\!4KK'' N'\!\!+\!K^2 N''')\!+\!FF'(4KK'N'\!+\!K^2 N'')\!-\!K^2N'(FF''\!-\! F'^2) \!+\!K^4 N'^3\Big),\nn\\ 
V_2&\!=\!&-V_5\!=\!-\tfrac{12 KF}{m^2 \ell^3}\Big(\!F^2(K'K''\!\!+\!\! \tfrac{KK'''}{3}\!)\!\!+\!\!K^2(3 KK'N'^2 \!\!+\!\!K^2 N' N''\!\!-\!\! F'F''\!\!-\!\! \tfrac{F F'''}{3}\! )\!\!+\!\!FF'(K'^2\!\!+\!\!KK'') \!-\!\! KK'(F'^2 \!\!-\!\!F'' F)\!\Big),\nn\\ 
V_3&\!=\!&-\tfrac{4K^2}{m^2 \ell^3}\Big(F^2(KK''N'\!-\!3K'^2 N'\!-\!KK' N'')\!+\!K^2(F F' N''\!-\! F F'' N'\!+\!K^2 N'^3\!-\! F'^2N') \!+\!4K K' F' FN' \Big),\\
V_4&\!=\!&\tfrac{4 K^2}{m^2 \ell^3}\!\Big(\!F^2(5 K K'' N' \!+\!9 K'^2 N'\!+\!7K K'N'')\!+\!2F F' (K^2N''\!\!+\!4 K K' N') \!+\! K^2(2K^2 N'^3\!-\!2 F F''N'\!\! -\!2 F'^2N'\!\!+\!F^2 N''')\Big).\nn
\eea

 \section{ENMG functions}
 The auxiliary fields $k^{a}$ and $h^a$ in the equations of motion of ENMG are defined as follows
\bea \label{kh} k^{0}=P_1 e^0+P_2e^2,\,\, k^{1}= P_3 e^1,\,\, k^{2}= P_4 e^2+ P_5 e^0,\qquad h^{0}=Y_1 e^0+ Y_2 e^2,\,\, h^{1}= Y_3 e^1,\,\, h^{2}= Y_4 e^2+Y_5 e^0,
\eea
where the functions $P_i$ and $Y_i$ are given by
\bea
Y_1&\!=\!&\frac{1}{8\ell^4}\Big(32 K^3 K' F^2 F' F''-40 K^5 K'' F^2N'^2 -24 K^6  F F''N'^2-32 K^2 K'K''F^3 F' +16 K^6 F^2 N''^2\nn\\
&\!\!\!+\!\!\!&32 K^3K'' F^2 F'^2+32  K^2K'^2 F^3 F''-96 KK'^2  K''F^4 +32 K^4 F F'' F'^2+32 K^3 K'F F'^3+9 K^8 N'^4\nn\\
&\!\!\!-\!\!\!&48 K^2 K''^2F^4 +16 K^4 F^2 F''^2+48 K^2 F^2K'^2 F'^2-24 K^6 F'^2N'^2-32 K K'^3F^3 F' -48 K'^4 F^4  \\
&\!\!\!+\!\!\!&216 K^4 K'^2 F^2N'^2+128 K^5K' F^2 N'  N''+16 K^4 F'^4+32 K^3 K''F^3 F''-24 K^5 K' F' F N'^2\Big)\nn\\
&\!\!\!+\!\!\!&\frac{m^2}{8\ell^2}\Big(16K^2 F F''-16K K''F^2 +16 K^2 F'^2-12 K^4 N'^2-16 K'^2 F^2+16 K K'F F' \Big),\nn
\eea
\bea
Y_2&\!=\!&-Y_5=\frac{ K^2 F}{\ell^4}\Big(12 K^4 K'N'^3 +3 K^5 N'^2 N''-4 K^2 K''F^2  N''-16 K K'^2F F' N'\nn\\
&\!-\!&4 K^2 K'F F' N''-4 K^3 F'^2 N''-16  K'^3 F^2N'-16 K^2 K'F'^2 N' -4 K^3 F F'' N'' \nn \\
&\!-\!&16 K K'K''F^2  N' -16 K^2 K'F F'' N'-4 K  K'^2F^2 N''\Big)-\frac{2m^2 K^2 F}{\ell^2}\Big(4K' N'+K N''\Big),\nn\\
Y_3&\!=\!&\frac{1}{8\ell^4}\Big(32 K^3 K'F^2 F'  F''+40 K^5 K''N'^2 F^2 -24 K^6  F F''N'^2+32 K^2 K' K''F^3 F' -16 K^6 F^2 N''^2\nn\\
&\!-\!&128 K^5 K' F^2 N' N''+32 K^4 F  F'^2F''+16 K^2 K''^2F^4+32 K K'^3F^3 F' +9 K^8 N'^4+16 K'^4 F^4\nn\\
&\!-\!& 24 K^6 F'^2 N'^2-32 K^3  K'' F^3F''-32 K^3 K''F^2  F'^2-32 K^2K'^2 F^3  F''+32 K K''K'^2 F^4 \nn \\
&\!+\!&32 K^3  K'F F'^3+16 K^4 F'^4 +16 K^4 F^2 F''^2-16 K^2 K'^2F^2 F'^2-216 K^4 K'^2 F^2N'^2 \nn\\
&\!-\!&24 K^5 K' F' F N'^2\Big)-\frac{m^2}{8\ell^2}\Big(16K^2 (F F''+F'^2)+16F^2(K K''+K'^2) -12 K^4 N'^2+16 K  K'F F'\Big),\nn\\
Y_4&\!\!=\!\!&-\frac{1}{8\ell^4}\Big(32 K^3 K'F^2 F'  F''+24 K^5 K'' F^2N'^2 -88 K^6  F F''N'^2 -16 K^6 F^2 N''^2-88 K^6  F'^2 N'^2\nn\\
&\!\!-\!\!&232 K^4 K'^2 F^2N'^2-48 K^2 K'^2F^2 F'^2+48 K^4 F^2 F''^2-40 K^5 K' F' F N'^2-32 K^2 K' K''F^3 F'\nn\\
&\!\!-\!\!&32 K^3 K''F^2  F'^2-32 K^2K'^2 F^3  F''-32 K K'^2 K'' F^4+96 K^4 F F'' F'^2-16 K^2K''^2 F^4 -16 K'^4 F^4 \nn\\
&\!\!-\!\!&128 K^5 K' F^2 N' N''-32 K^3 K''F^3  F''+48 K^4 F'^4 +32 K^3 K'F F'^3-32 K K'^3F^3 F'+39 K^8 N'^4\Big)\nn\\
&\!-\!&\frac{m^2}{8\ell^2}\Big(16K^2 (F F''+F'^2)-16F^2( K K''+K'^2) -20 K^4 N'^2-16 K K'F F' \Big),\nn
\eea
and
\bea
P_1&\!=\!&\frac{2 K^2}{\ell^5}\Big(7K^6 F'^2 N'^3\!-\!56 K^4 K'^2 F^2  N'^3\!+\!12K^3K'FF'^3N'\!+\!48K'^4 F^4 N'\!+\!80 K^2K'K'' F^3 F' N'\!-\!3K^8 N'^5\nn\\
&\!+\!&16K^2 K''^2 F^4 N'\!-\!4 K^4  F^2 F''^2N'\!+\!32 KK'^3F^4 N''\!+\!4 K^3K'''F^4  N''\!+\!4 K^4 F^3 F'' N'''\!+\!4 K^4 F^2 F'^2 N'''\nn\\
&\!+\!&4 K^3 K''F^4  N'''\!+\!4K^2K'^2 F^4  N'''\!+\!116 K^2 K'^2 F^2 F'^2 N'\!+\!100K^2  K'^2 F^3 F''N'\!-\!3 K^6 F^2 N'^2 N'''\nn\\
&\!+\!&7K^6  F F''N'^3\!+\!16 K^3 K''F^3 F'  N''\!+\!16 K^4F^2 F' F'' N''\!-\!16 K^5 K'' F^2 N'^3\!-\!4 K^6 F^2 N' N''\!+\!4K^4 F^3 F'''N'' \nn\\
&\!-\!&8 K^4  F F'' F'^2N'\!-\!8K^5K'FF'N'^3\!+\!112 K  K'' K'^2F^4 N'\!+\!112 K  F F'' F'^2N'\!+\!112KK'^3 F^3 F'N'\\
&\!+\!&44 K^3  K' F^3F'' N''\!+\!20 K^3 K''F^3 F'' N'\!-\!3K^6 F F' N'^2 N''\!+\!16 K^2K'K''' F^4  N' \!+\!4 K^3 K'F^3 F'  N'''\nn\\
&\!-\!&4 K^4  F'^4N'\!-\!8 K K''F^2 N'\!+\!20 K^3K'' F^2 F'^2 N'\!+\!48 K^2 K'^2F^3  F' N''\!+\!60 K^3  K'F^2  F' F''N'\nn\\
&\!+\!&4K^4 F F'^3 N''\!+\!48 K^3K'  F^2 F'^2 N''\!+\!16 K^3 K'F^3 F''' N' \!-\!44 K^5 K' F^2 N'^2 N''\!+\!44 K^2 K' K'' F^4 N''\Big)\nn\\
&\!+\!&\frac{2 m^2 K^2}{\ell^3}\Big(8 KK'FF'N'+24 K'^2F^2 N'\!+\!2K^2 F^2N'''\!+\!2K^4N'^3\!+\!16KK'F^2N''\nn\\
&\!+\!&2K^2FF'N''\!+\!8KK''F^2N'\!-\!2K^2FF''N'\!-\!2K^2 F'^2N'\Big),\nn
\eea
\bea
P_2&\!=\!&\frac{2 KF}{\ell^5}\Big(12 K'^4 F^3 F'\!-\!36K^3 K'F'^4\! +\!12 K'^3 K''F^4\!-\!8 K^2 K'^2F F'^3 \!+\!32 K K'^3F'^2 F^2 \!-\!51 K^7  K'N'^4 \!-\!36 K^4F'^3  F''\nn\\
&\!+\!&93K^5 K' F'' FN'^2 \!+\!180 K^3  K'^3 F^2N'^2\!+\!16 K^5 K' F^2 N' N'''\!\!-\!K^5  K'' F' FN'^2\!+\!12 K^2 K''^2F^3  F'\!-\!12 K^4 F^2 F'' F'''\nn\\
&\!+\!&43 K^4  K'K'' F^2 N'^2\!+\!20 KK'^3 F^3 F''\!+\!12 K^5 K'' F^2 N' N''\!+\!24 K^5  F F'' N'N''\!-\!36 K^3 K' F^2 F''^2\!+\!4 K^2  K'' K'''F^4\nn \\
&\!+\!&12 K K' K''^2 F^4\!+\!172 K^4  K'^2 F^2 N'N''\!+\!4 K^3 K''F^3  F'''\!+\!4 K^3  K'' F^3F'''\!+\!4K^3K''' F^3  F'^2\!+\!44 K^5 K' F F' N' N''\nn\\
&\!+\!&4 K^2K'^2 F^3  F'''\!-\!12 K^4 F F'^2F'''\!+\!16K^3 K''F^2 F'  F''\!-\!21 K^8 N'^3 N''\!+\!4 K  K'^2 K'''F^4\!+\!99 K^4 K'^2 FF'N'^2 \nn \\
&\!\!\!\!+\!&11 K^6 F F'''N'^2 \!+\!4 K^2 K' K'''F^3  F'\!+\!32 K^2 K'K'' F^3  F''\!+\!44K^2 K' K''F'^2 F^2\!+\!4  K^3 K''F F'^3 \!-\!4  K^3 K' F^2 F'F'''\nn\\
&\!+\!&4 K^6 F^2 N'' N'''\!+\!4 K^2 K'^2F^2 F'  F'' \!+\!32 K^5 K'F^2 N''^2\!+\!36 K K'^2 K''F^3 F'\!-\!36 K^4 F F''^2 F' \!+\!93 K^5 K'F'^2 N'^2\nn\\
&\!+\!&33 K^6 F' F'' N'^2\!+\!24 K^6 F'^2 N' N''\!-\!84 K^3 K' F F'^2 F''\!+\!4 K^6 F  F'N'^2\!-\!\,3 K^5 K'''F^2 N'^2 \Big)\!+\!\frac{2 m^2 KF}{\ell^3}\Big(18K^3 K' N'^2\nn\\
&\!+\!&6 K^4 N' N'' \!-\!6K K' F'^2\!+\!6K'^2 F F'\!+\!6 K' K'' F^2\!-\!6 K^2 F' F''\!-\!2K^2 F F'^3\!+\!2 K  K'''F^2\!-\!6 K K'F F''\!+\!6  KK''F F'\Big),\nn
\eea
\bea
P_3&\!=\!&\frac{2 K^2}{\ell^5}\!\Big(\!7 K^6  F'^2N'^3\!+\!13 K^4  K'^2 F^2N'^3\!+\!12 K^3 K' F F'^3 N'\!+\!4K K'^3 F' F^3 N'\!+\!20 K^2 K' K''F^3 F' N'\nn\\
&\!+\!&4 K^2 K''^2 F^4 N'\!-\!4 K^4  F^2 F''^2N'\!-\!4 K  K'^3F^4 N''\!+\!4 K^4 F F'^3 N''\!+\!4 K^4 F^2 F' F'' N''\!+\!K^5 K'' F^2 N'^3 \nn\\
&\!+\!&4K^3 K''F^3 F'  N'' \!+\!7K^6 F F''N'^3\!-\!8 K K'^2 K''F^4 N' \!-\!8 K^5 K' F' F N'^3\!-\!8 K^4  F F'' F'^2N'\nn\\
&\!-\!&4 K^3 K' F^3 F'' N''\!+\!3K^5 K' F^2  N'^2N''\!-\!3 K^6 F F' N'^2 N''\!-\!3 K^8 N'^5\!-\!12K'^4 F^4 N'\!-\!4 K^4 F'^4N' \nn\\
&\!-\!&16 K^2  K'^2 F^3 F''N'\!+\!12 K^3 K'F^2 F' F''N'\!-\!4 K^2K' K''  F^4 N''\Big)-\frac{2 m^2 K^2}{\ell^3}\Big(2 KF^2(K'  N''\!-\!K'' N')\nn\\
&\!+\!&2K^2 (F'^2N'\!+\!F F''N'\!-\!F F' N'' ) \!+\!K' F N'(6  K' F\!-\!8 K F') \!-\!2K^4 N'^3 \Big),\nn
\eea
\bea
P_4&\!=\!&\frac{2 K^2}{\ell^5}\!\Big(\!14 K^6  F'^2N'^3\!-\!43 K^4  K'^2 F^2N'^3\!+\!24 K^3 K' F F'^3N' \!+\!116 K K'^3 F' F^3N' \!-\!16 K^5 K' F' FN'^3\nn\\
&\!-\!&6 K^8 N'^5\!-\!3 K^6 F^2 N'^2 N'''\!+\!36 K'^4 F^4N'\!+\!20 K^2 K''^2 F^4 N'\!-\!8K^4  F^2 F''^2N'\!+\!28 K  K'^3F^4 N''\nn\\
&\!+\!&4 K^2K'^2 F^4  N'''\!+\!4 K^4 F^3 F''' N''\!+\!4 K^4 F^3 F''N'''\!+\!4 K^4 F^2 F'^2 N'''\!+\!4K^3  K'''F^4 N''\!+\!8 K^4 F F'^3 N''\nn\\
&\!+\!&20 K^3 K''F^3 F'  N''\!+\!14 K^6  F F''N'^3\!+\!40 K^3 K'F^3  F'' N''\!+\!84 K^2  K'^2 F^3 F''N' \!+\!20 K^4 F^2 F' F'' N''\nn \\
&\!-\!&16 K^4 K'' F^2N' \!-\!4 K^6 F^2 N' N''\!-\!15 K^5K'' N'^3 F^2\!+\!72 K^3  K' F^2 F' F''N'\!+\!4 K^3 K'F^3 F'  N'''\!-\! 8 K^4  F'^4N'\nn\\
&\!-\!&6 K^6 F F' N'^2 N''\!+\!16 K^2 K'K'''F^4  N' \!+\!16 K^3 K'F^3 F''' N' \!+\!20 K^3 K''F^3 F'' N' \!+\!20 K^3 K''F^2 F'^2 N'\nn\\
&\!+\!&48 K^2 K'^2F^3  F' N''\!+\!48 K^3 K'F^2  F'^2 N''\!+\!4 K^3 K''F^4  N'''\!+\!40 K^2 K' K''F^4  N'\!-\!41 K^5 K'F^2 N'^2  N''\nn\\
&\!+\!&116 K^2 K'^2F^2  F'^2 N'\!+\!100 K^2 K' K'' F^3 F' N'\!+\!104 K  K'^2K'' F^4 N' \Big)\! -\!\frac{2 m^2 K^2}{\ell^3}\Big(4K^2  F'^2N'\!-\!2 K^2 F^2 N'''\nn\\
&\!-\!&4K^4 N'^3\!-\!4 K^2 F F' N''\!-\!10 K K'' F^2 N'\!-\!18  K'^2 F^2N' \!-\!16 K K' F' F N'\!-\!14 K K'F^2  N''\!+\!4 K^2  F F''N'\Big),\nn
\eea
\bea
P_5&\!=\!&-\frac{2 KF}{\ell^5}\Big(36 K'^4 F^3 F'\!-\!4 K^2 K' K''F'^2 F^2 \!-\!69 K^4 K'^2 FF'N'^2 \!-\!32 K^2 K'^2F F'^3  \!-\!128 K^4  K'^2 F^2N' N''\nn\\
&\!+\!&45 K^5 K' F'^2 N'^2\!+\!45 K^5  K'N'^2 FF''\!-\!84 K^3  K'^3 F^2N'^2 \!+\!8 KK'^3 F^2  F'^2\!+\!36 K^2 K''^2F^3  F'' \!-\!12  K^4F'^3 F''\nn\\
&\!+\!&23 K^5 K'' FF'  N'^2\!-\!20 K^3 K''F F'^3 \!-\!4  K K'^3F^3 F'\!-\!5 K^4  K' K''F^2 N'^2\!+\!12 K^6  F F''N' N''\!-\!12 K^3 K'F'^4\nn\\
&\!+\!&36 k K' K''^2 F^4\!+\!12 K^2 K'' K'''F^4 \!-\!12 K^4 F  F'F''^2\!-\!12 K^3 K'F^2 F''^2\!-\!4 K^4 F^2 F'' F'''\!-\!52 K^5K'  F' F N' N'' \nn\\
&\!-\!&4 K^3 K'''F^2  F'^2\!-\!4 K^2K'^2 F^3  F'''\!-\!4 K^3 K''' F^3 F''\!-\!4 K^3  K''F^3 F'''\!-\!32 K^3 K'' F^2 F' F''\!+\!12 K  K''' K'^2F^4\nn\\
&\!-\!&4 K^3 K'F^2 F'  F'''\!-\!4 K^4 F  F'^2F'''\!+\!36 K'^3 F^4 K''\!-\!44 K^2 K'^2 F^2 F' F''\!+\!5 K^5 K''' F^2 N'^2\!-\!16 K^5 K'F^2 N'  N''' \nn\\
&\!+\!&3 K^6  F F'''N'^2\!+\!4 K^2K' K'''F^3 F'  \!+\!12 K^6  F'^2N' N''\!-\!16 K^2 K' K'' F^3F''\!-\!36 K^3 K'F F'^2  F'' \!-\!8 K^6 F F'N''^2\nn\\
&\!-\!&4 K^6 F^2 N'' N'''\!-\!27 K^7  K'N'^4 \!-\!9 K^8 N'^3 N''\!+\!9K^6  F' F''N'^2 \!-\!28 K^5 K'F^2 N''^2\!+\!84  K K'^2 K''F^3F'\Big)\nn\\
&\!+\!&\frac{2 m^2 KF}{\ell^3}\Big(6 K( K'F'^2\!+\!K F' F''\!-\!K^3 N' N''\!)\!-\!6F(K'(FK''\!+\!K F')\!+\!K(K' F''\!-\!K''F'))\!+\!2K F(KF'''\!-\!K'''^2)  \Big).\nn
\eea
 

\end{appendix}

\end{document}